\documentclass[aps, prl, twocolumn]{revtex4-1}
\usepackage{graphicx}
\usepackage{float}
\usepackage{amsmath}
\usepackage{amsfonts}
\usepackage{color}

\begin{document}

\title{Deconstructing Magnetization Noise: Degeneracies, Phases, and Mobile Fractionalized Excitations in Tetris Artificial Spin Ice}

\author{M. Goryca$^{1,2}$, X. Zhang$^3$, J. Ramberger$^{4}$, J. D. Watts$^{4,5}$, C. Nisoli$^6$, C. Leighton$^4$, P. Schiffer$^{3,7}$, S. A. Crooker$^1$}
\affiliation{$^{1}$National High Magnetic Field Lab, Los Alamos National Laboratory, Los Alamos, NM 87545, USA}
\affiliation{$^{2}$Institute of Experimental Physics, Faculty of Physics, University of Warsaw, Pasteura 5, 02-093 Warsaw, Poland}
\affiliation{$^{3}$Department of Applied Physics, Yale University, New Haven, CT 06520, USA}
\affiliation{$^{4}$Department of Chemical Engineering and Materials Science, University of Minnesota, Minneapolis, MN 55455, USA}
\affiliation{$^{5}$School of Physics and Astronomy, University of Minnesota, Minneapolis, MN 55455, USA}
\affiliation{$^{6}$Theoretical Division, Los Alamos National Laboratory, Los Alamos, NM 87545, USA}
\affiliation{$^{7}$Department of Physics, Yale University, New Haven, CT 06520, USA}

\begin{abstract}
Direct detection of spontaneous spin fluctuations, or ``magnetization noise'', is emerging as a powerful means of revealing and studying  magnetic excitations in both natural and artificial frustrated magnets. Depending on the lattice  and nature of the frustration, these excitations can often be described as fractionalized quasiparticles possessing an effective magnetic charge. Here, by combining ultrasensitive optical detection of thermodynamic magnetization noise with Monte Carlo simulations, we reveal emergent regimes of magnetic excitations in artificial ``tetris ice''. A marked increase of the intrinsic noise at certain applied magnetic fields heralds the spontaneous proliferation of fractionalized excitations, which can diffuse independently, without cost in energy, along specific quasi-1D spin chains in the tetris ice lattice. 
\end{abstract}

\maketitle

While the existence of elementary magnetic monopoles remains hypothetical, spin (i.e., dipole) excitations in certain frustrated magnetic systems \cite{Bramwell:2020, Skjaervo:2020, Rougemaille:2019, Nisoli:2021} can fractionalize and separate into two delocalized ``monopole-like'' quasiparticles that each carry an effective magnetic charge \cite{Castelnovo:2008, Mol:2009, Morris:2009, Jaubert:2009, Ladak:2010, Castelnovo:2012, Perrin:2016, Farhan:2019}. These fractionalized excitations are topologically protected, can diffuse through the crystal lattice in thermal equilibrium, and can move in response to applied magnetic fields, motivating studies of ``magnetricity'' in analogy to electricity \cite{Bramwell:2009, Giblin:2011, Pollard:2012, Vedmedenko:2016, Morley:2019}. 

Mobile magnetic quasiparticles have been investigated in both natural spin ice materials (such as the 3D pyrochlore Dy$_2$Ti$_2$O$_7$ \cite{Bramwell:2009}) and also in engineered 2D arrays of nanomagnets known as artificial spin ice (ASI) \cite{Ladak:2010, Mengotti:2011, Morgan:2011, Phatak:2011}. Recently, powerful new detection modalities have emerged for revealing and studying these fractionalized excitations, based on direct sensing of the very small -- but measurable -- magnetization fluctuations that arise from their stochastic creation, annihilation, and diffusion in thermal equilibrium. In Dy$_2$Ti$_2$O$_7$, spontaneous magnetization noise was detected at cryogenic temperatures via SQUID magnetometry \cite{Dusad:2019, Samarakoon:2022, Hallen:2022}, from which the presence of monopoles and their dynamics were inferred.  Separately, in archetypal square ASI lattices, magnetization noise from quasiparticle kinetics was detected at room temperature via optical magnetometry \cite{Goryca:2021, Goryca:2022}. Crucially, in these latter studies, the sudden appearance of excess noise at certain applied magnetic fields revealed the presence of phases rich in mobile magnetic charges. 

Motivated by these studies, here we use optical magnetometry of spontaneous noise to reveal new families of fractionalized  excitations in the low-symmetry frustrated ASI known as ``tetris ice'' \cite{Morrison:2013, Gilbert:2016}. By applying small magnetic fields, tetris ice can be tuned through a variety of complex spin configurations, whose degeneracies and boundaries are directly revealed by noise. In particular, we find particularly intense and narrow bands of noise for certain directions and ranges of applied field.  Using Monte Carlo simulations to deconstruct these noise signatures, these bands are shown to herald novel regimes wherein magnetic quasiparticles delocalize and proliferate, without cost in energy, along extended quasi-1D spin chains in the lattice.   These results demonstrate the power of noise-based studies to probe microscopic details of complex magnetic phenomena, in this case specifically the equilibrium kinetics driven by fractionalized magnetic excitations. 

Figure 1a shows a scanning electron microscope (SEM) image and illustration of tetris ice. The individual islands have lateral dimensions 220 $\times$ 80~nm and are made of ferromagnetic Ni$_{0.8}$Fe$_{0.2}$ using established lithographic methods \cite{Gilbert:2016}. Each island behaves as a single Ising-like macrospin, with magnetization orientation parallel or antiparallel to its long axis. Crucially, the islands are very thin ($\approx$3.5 nm), so that at room temperature they behave as thermally-active superparamagnets (i.e., their magnetization direction thermally fluctuates in the absence of a strong biasing magnetic field \cite{Kapaklis:2014, Chen:2019}). This ensures that the ASI can efficiently explore the huge manifold of possible magnetic configurations, and remain at or near its lowest energy in thermal equilibrium. 

Recent studies of tetris ice demonstrated that its low-energy configuration at zero applied field ($B$=0) comprises ordered ``backbones'' (blue islands in Fig. 1a) that include the four-fold coordinated vertices ($z$=4), separated by disordered ``staircases'' (red islands) containing $z$=3 and $z$=2 vertices \cite{Gilbert:2016}. At $B$=0, the phenomenon of vertex frustration \cite{Morrison:2013} prevents all the staircase vertices from achieving their lowest-energy configuration, leading to extensive degeneracy \cite{Gilbert:2016} and entropy-driven ordering \cite{Saglam:2022, Miller:2022}.

Despite the intriguing physics observed in tetris ice at $B$=0, its properties have never been explored in applied magnetic fields. However, given its lower symmetry and the demonstrated emergence of complex collective behavior, a rich set of phenomena in the presence of magnetic fields can be anticipated.  We investigate the field-dependent behavior of tetris ice by optically detecting its intrinsic magnetization noise, using the approach shown in Fig. 1b. This method is adapted from earlier studies of optically detected spin noise in atomic, semiconductor, and ferromagnetic systems \cite{Crooker:2004, Crooker:2010, Zapasskii:2013, Balk:2018}. Here, spontaneous magnetization fluctuations in thermal equilibrium are passively detected via the Kerr rotation fluctuations that they impart on a linearly-polarized laser reflected from the ASI surface. These fluctuations are detected in real time by balanced photodiodes, and the total noise power is computed via fast-Fourier transform methods. Small coils apply magnetic fields $B_x$ and $B_y$ in the sample plane. By reorienting the laser with respect to the sample, we measure fluctuations of either the horizontal islands, $\langle (\delta m_x)^2 \rangle$, or the vertical islands, $\langle (\delta m_y)^2 \rangle$. 

Figures 1c and 1d show the main experimental results, which are field-dependent maps of the spontaneous magnetization noise from the horizontal and vertical islands in tetris ice. The maps reveal a complex structure, with both sharp and diffuse bands of noise appearing along certain directions and regions of $B_x$ and $B_y$. The dark areas, where noise is absent, indicate regions of stable (non-fluctuating) magnetic order, which are separated by noisy boundaries. As discussed below in Fig. 2, these boundaries reveal where different magnetic configurations become energetically degenerate. Note that these maps are not related by 90$^{\circ}$ rotation, reflecting the absence of $C_4$ symmetry. Fig. 1d even shows an annular region of noise surrounding $B \approx 0$ (but Fig. 1c does not), which is related to fluctuations within the backbones and will be discussed later.   

Of primary interest, and the main focus of this work, are the bright and narrow bands in the noise map where spontaneous fluctuations are especially strong (labeled A and B, indicated by dashed lines). These bands appear only for sufficiently large applied fields, and follow specific diagonal trajectories in the $(B_x, B_y)$ plane.  Motivated by recent studies of archetypal square ASI, wherein the sudden onset of spontaneous noise revealed regimes of highly mobile monopoles, we therefore seek to identify the origin of these intense fluctuations in tetris ice. 

\begin{figure}[t]
\center
\includegraphics[width=.45\textwidth]{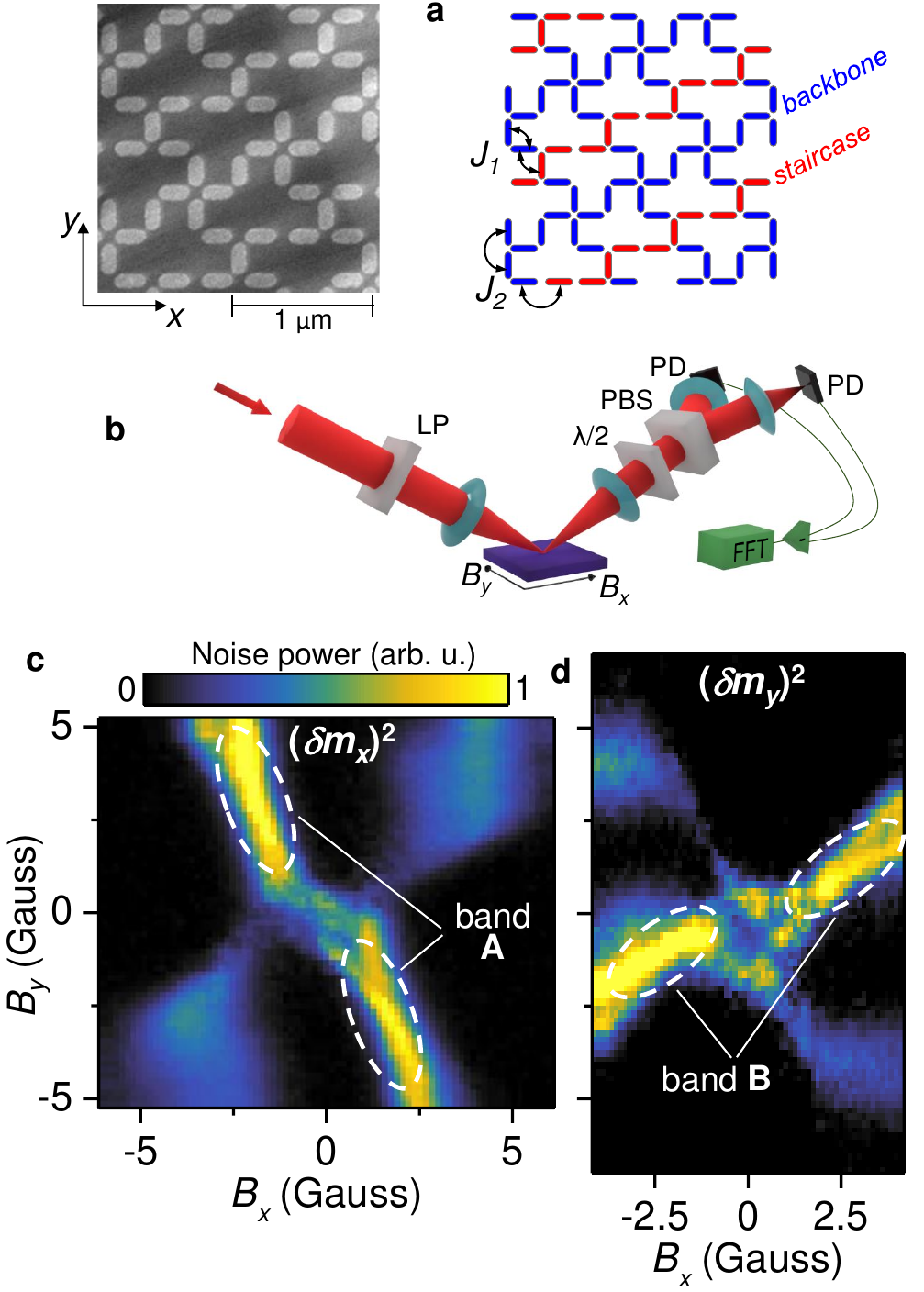}
\caption{a) SEM image and diagram of tetris ice. The lattice comprises `backbones' (blue islands) and `staircases' (red islands). b) Experiment: Magnetization fluctuations are passively detected via the polarization (Kerr rotation) fluctuations that they impart on a linearly-polarized laser reflected from the ASI. c) A map of the measured noise power from the horizontal islands, $\langle (\delta m_x)^2 \rangle$, in thermal equilibrium, vs. in-plane magnetic fields $B_x$ and $B_y$.  d) Same, but for the vertical islands, $\langle (\delta m_y)^2 \rangle$.  Intense noise along bands A and B herald the proliferation of fractionalized magnetic excitations.}
\label{Fig1}
\end{figure}

Crucial insight into these complex noise maps is provided by Monte Carlo (MC) simulations of Glauber spin dynamics, following our earlier simulations of noise in square and quadrupolar ASI \cite{Goryca:2021, Goryca:2022}. We simulated 32$\times$32 vertex tetris lattices with periodic boundary conditions, and only nearest- and next-nearest neighbor interactions were considered ($J_1$ and $J_2$; see Fig. 1a). For each MC time step, $N$ single islands were chosen randomly ($N$ is the number of islands in the lattice; cluster and loop flips were not considered), and the spin flip acceptance probability was $[1 + \mathrm{exp}(\Delta/kT)]^{-1}$, where $\Delta$ is the energy difference associated with the spin flip and $k$ is Boltzmann's constant.  We used $J_1 = 1.7 J_2$, and temperature $kT = 0.7 J_2$ that corresponds to room temperature in our structures; both values are consistent with micromagnetic MuMax3 simulations \cite{MuMax3} of these permalloy islands \cite{Goryca:2021}. These values were also validated by directly comparing measured and simulated noise maps. In particular, as discussed below, the positions of the various noise bands is determined by $J_1$ and $J_2$, and their widths are determined in part by $kT$. At each value of ($B_x$, $B_y$), $10^6$ MC annealing steps were performed to ensure thermal equilibrium, and then the magnetization was recorded for several million additional MC time steps. The average magnetization $M$ and the thermodynamic fluctuations $\langle [\delta M(t)]^2 \rangle$ about this mean value were determined from the computed time series.


Figure 2 shows a calculated map of the field-dependent magnetization, revealing a rich tableau of magnetic configurations (color and brightness indicate the direction and magnitude of \textbf{M}). Dotted black lines denote boundaries between the different static configurations that are stabilized at high field when $B_x$ and/or $B_y$ are large (labelled \textit{i-xii}, and depicted around the periphery of the map). Beginning in the upper-right corner of the map where both $B_x$ and $B_y$ are large and positive (region \textit{i}), all horizontal and vertical islands are trivially magnetized along $+\hat{x}$ and $+\hat{y}$, respectively, by the applied field. Proceeding clockwise around the map (\textit{i.e.}, as $B_y \rightarrow 0$), a subset of the vertical islands flips to $-\hat{y}$ when $B_y<J_1 / \mu$, where $\mu$ is the moment of an island; that is, when the Zeeman energy balances the interaction energy, and the lattice transitions to configuration \textit{ii}. The pink arrows in the diagram next to region $ii$ show which islands have flipped upon transitioning from $i \rightarrow ii$: here it is only the subset of vertical islands in the $z$=4 vertices; these have an `unbalanced' number of horizontal neighbors. Continuing clockwise around the perimeter, from region $ii \rightarrow iii \rightarrow iv \rightarrow v~...$, different subsets of islands (pink arrows) flip when crossing each boundary. 

\begin{figure}[t] 
\center
\includegraphics[width=.45\textwidth]{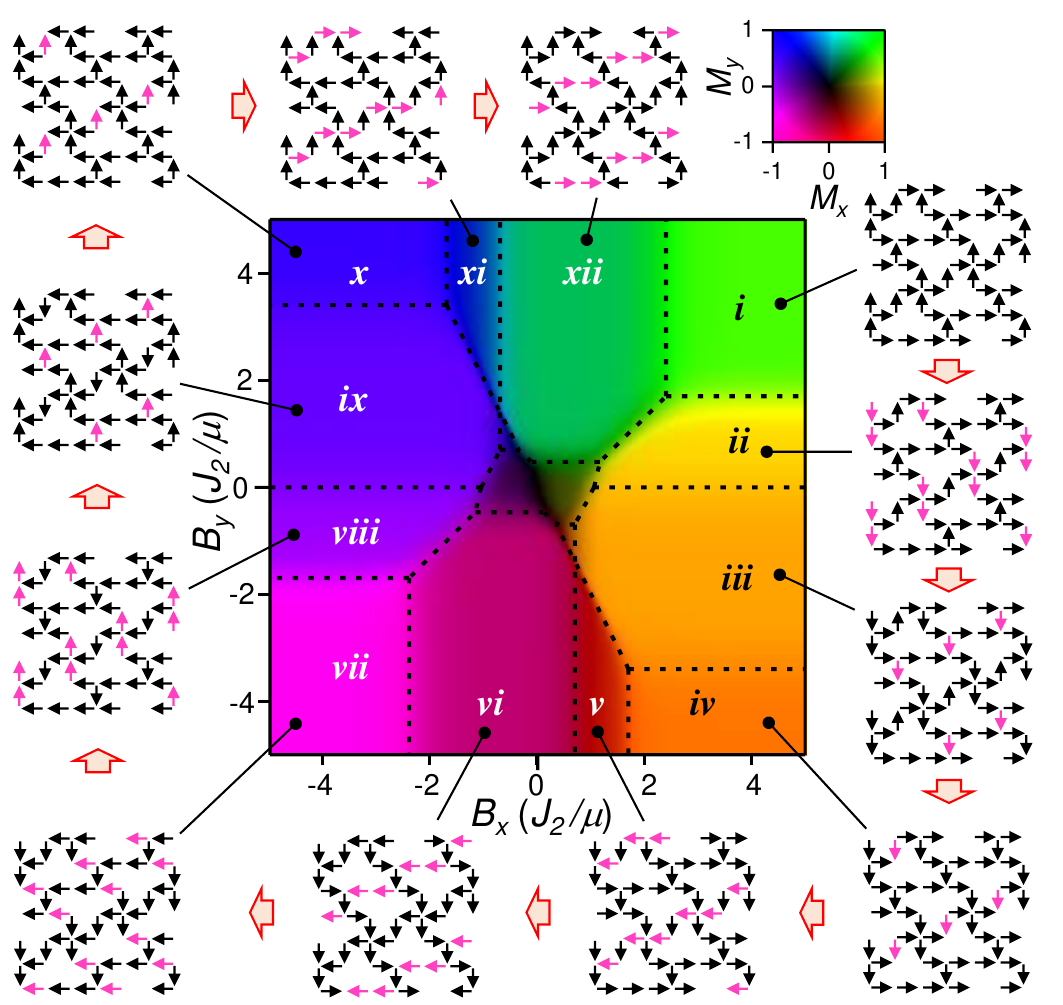}
\caption{Magnetic states of tetris ice, vs. $B_x$ and $B_y$. The color and brightness indicate the direction and magnitude of the average magnetization \textbf{M}, as computed by MC simulations. Dotted lines denote boundaries between different magnetic configurations (labeled \textit{i-xii}) that are stabilized when $|B_{x,y}|$ is large. Moment configurations are shown; pink arrows indicate the subset of islands that flip when transitioning from state $i\rightarrow ii \rightarrow iii$~..., clockwise around the map's perimeter. Boundaries occur when Zeeman energies balance the local interaction energies given by $J_{1,2}$. The dark central region is discussed in the Supplemental Material.}
\label{fig2}
\end{figure}

From this diagram it is already possible to associate the experimentally-measured bands of noise with boundaries between stable magnetic configurations, and more importantly that the bands A and B of exceptionally high noise occur at the internal diagonal $iii \leftrightarrow v$ boundary (or equivalently, $ix \leftrightarrow xi$), and at the diagonal $xii \leftrightarrow ii$ ($vi \leftrightarrow viii$) boundary.

\begin{figure}[t] 
\center
\includegraphics[width=.45\textwidth]{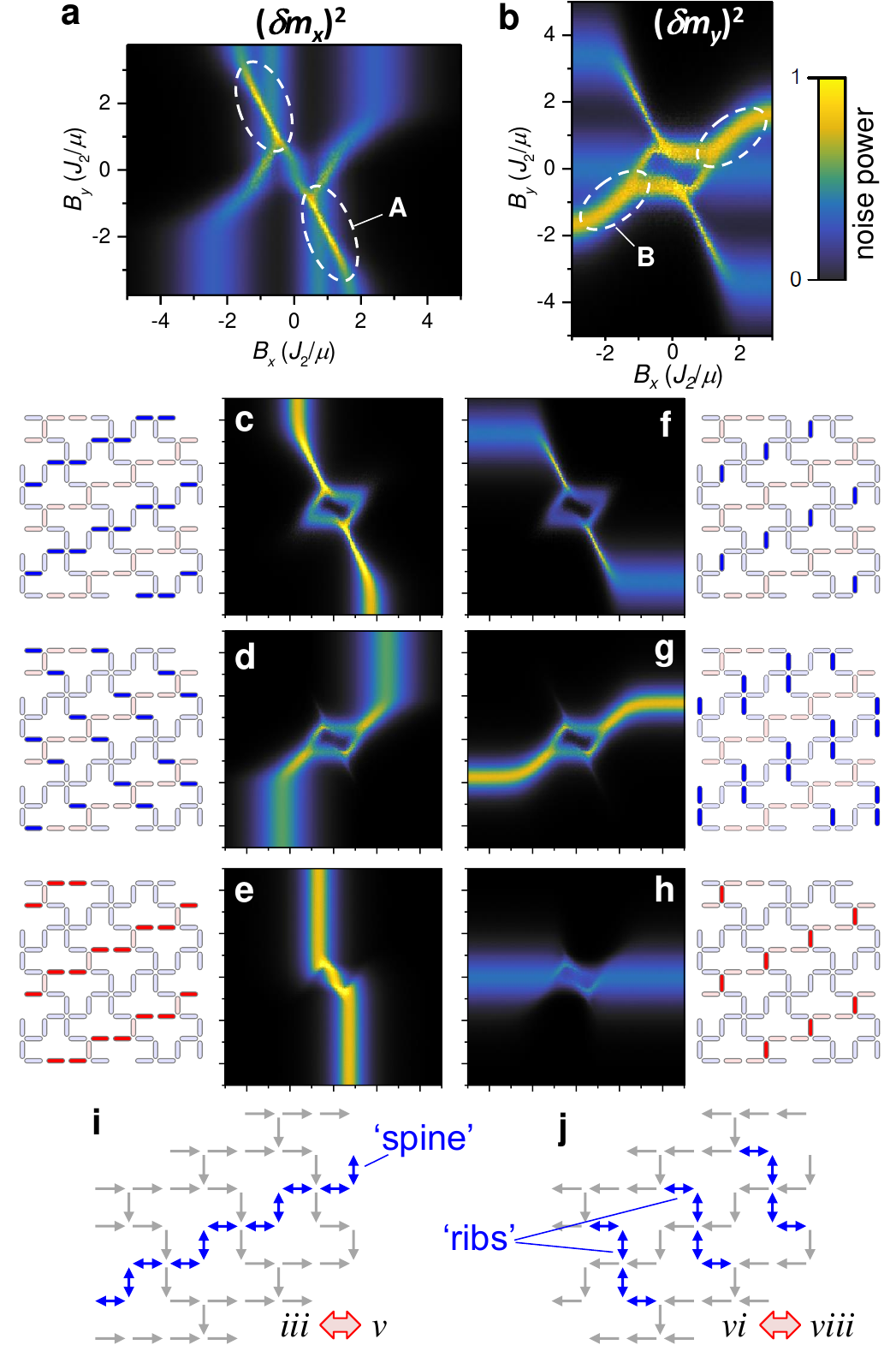}
\caption{Calculated maps of the equilibrium magnetization noise from a) the horizontal and b) the vertical islands, showing excellent agreement with experiment (\textit{cf.} Fig. 1). c-e) Deconstructing noise: Noise maps from \textit{specific subsets} of horizontal islands, indicated by bright colors (blue and red denote backbone and staircase islands). f-h) Same, but for subsets of the vertical islands. i, j) Blue arrows indicate the 1D spin chains that flip when crossing the diagonal $iii \leftrightarrow v$ and $vi \leftrightarrow viii$ boundaries, respectively.  These chains form the ``spines'' and ``ribs'' of the backbones. Along these boundaries, the balance of Zeeman and interaction energies restores the Z2 symmetry of the spin chains.}
\label{fig_square_deg} 
\end{figure}

This association is confirmed in Figs. 3a and 3b, which show the calculated noise power from the horizontal and vertical islands [$(\delta m_x)^2$ and $(\delta m_y)^2$, respectively]. Crucially, the overall agreement with the experimental data  is remarkably good (\textit{cf.} Fig. 1), including capturing both narrow and diffuse bands of noise, as well as the small annulus of noise surrounding $B \approx 0$ in $(\delta m_y)^2$. Because the MC simulations are validated in this way, we have confidence that they can be explored in detail to reveal the correlated physics of the system. Most importantly, the simulations allow us to determine \textit{which} subset(s) of islands give rise to the different noise signatures that are observed experimentally.  In other words, we can use the validated MC simulations to deconstruct the noise maps and tease apart the fluctuations of individual moments in ways that are experimentally intractable.  

To this end, Figs. 3(c-e) show the calculated noise from specific subsets of the horizontal islands (indicated by bright blue and red in the adjacent diagrams). Similarly, Figs. 3(f-h) show noise from subsets of vertical islands. From these noise maps it is clear that the horizontal and vertical bands of noise appearing at large $B_{x,y}$ originate from subsets of disconnected and uncorrelated islands that become thermally active when crossing a horizontal or vertical configuration boundary in Fig. 2. 

\begin{figure} 
\center
\includegraphics[width=.48\textwidth]{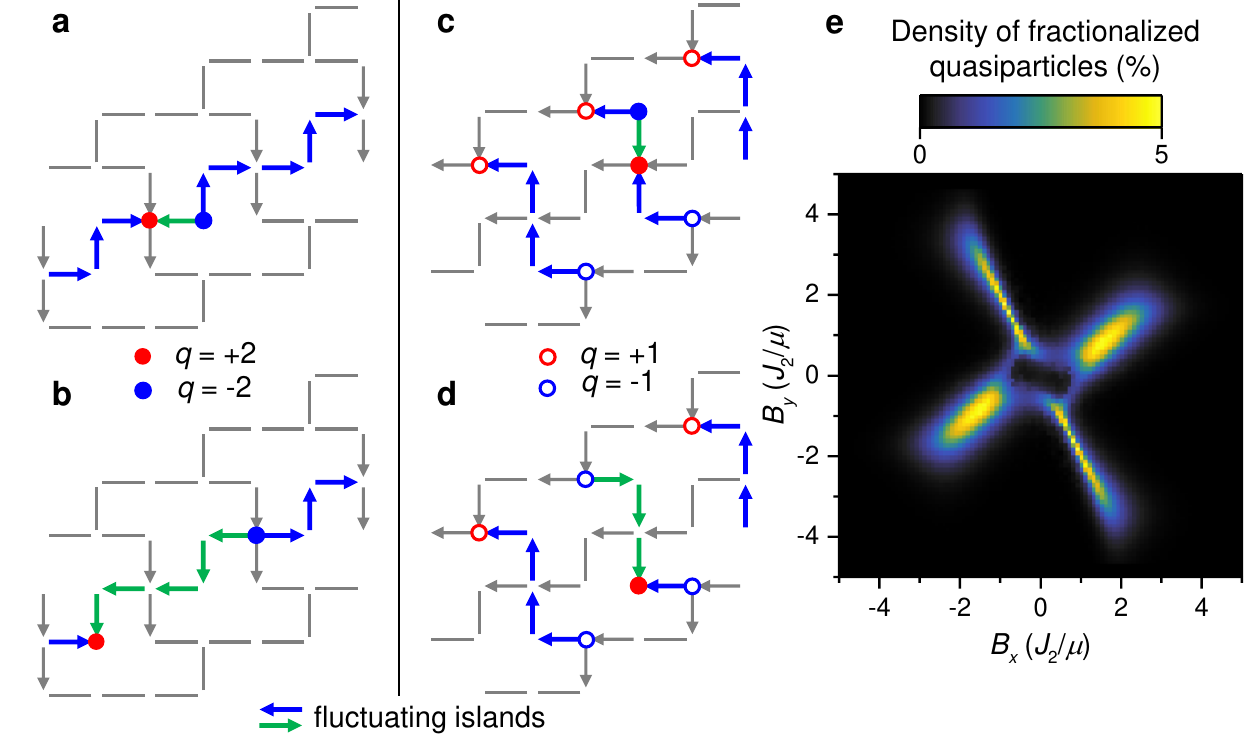}
\caption{a) A thermal spin-flip of one island (green) in a ordered 1D ``spine'' generates a dipole excitation and two adjacent vertices with magnetic charge $q=\pm 2$. b) Subsequent flips of neighboring islands fractionalize the excitation into two delocalized quasiparticles that move along the spine, leaving behind a string of flipped islands. For $(B_x,B_y)$ along the $iii \leftrightarrow v$ boundary this motion costs no net energy, and the charges therefore diffuse independently and proliferate, leading to the excess noise observed in band A. If $(B_x,B_y)$ is tuned off the boundary, the $\pm$2 charges move preferentially in opposite directions and annihilate. c,d) Similar phenomena along the ``ribs'', at the $vi \leftrightarrow viii$ boundary (band B). Note that ordered ribs end in $z$=3 vertices, with $q = \pm$1; therefore excitations have $q= \mp$1 or $\pm 2$. e) Map of the density of fractionalized (delocalized) quasiparticles.}
\label{fig4}
\end{figure}

Most importantly, these deconstructed maps reveal the origins of the intense noise that emerges in the diagonal bands A and B of the experimental data.  Figures 3c and 3f, considered together, show that band A arises from fluctuations of the horizontal \textit{and} vertical islands comprising the extended 1D spin chains that connect the $z=4$ vertices (\textit{i.e.}, the ``spines'' of the backbones). As Fig. 3i shows, it is precisely and only these connected spin chains that flip at the diagonal $iii \leftrightarrow v$ boundary (and $ix \leftrightarrow xi$, by symmetry). Along this border, configurations \textit{iii} and \textit{v} are energetically degenerate and the balance of Zeeman and interaction energies restores the Z2 spin symmetry of the entire chain.  Similarly, Figs. 3d and 3g show that noise in band B arises from the four-island chains that form the ``ribs'' of the backbones; only these spin chains are degenerate at the diagonal $ii \leftrightarrow xii$ (and $vi \leftrightarrow viii$) boundary (see Fig. 3j).  

Figures 4a,b show how fluctuations at the $iii \leftrightarrow v$ boundary arise from spontaneous creation, fractionalization, and kinetics of magnetic quasiparticles along the spin chains that form the spines of the backbones. Vertices along these chains have coordination \textit{z}=2 or \textit{z}=4, and an effective magnetic charge, $q$, given by the balance of inward- and outward-facing moments. In region \textit{iii}, all islands in the chain have stable orientation up or right, and all vertices have $q=0$ (2-in, 2-out at the $z$=4 vertices, and 1-in, 1-out at the $z$=2 vertices).  Just across the border, in region \textit{v}, all islands in the chain have orientation down or left, and again all vertices have $q=0$. Along the border, however, the two configurations are degenerate and Z2 symmetry of the spin chain is restored. Beginning with an ordered chain, a thermal spin-flip of any island in the chain creates a dipole excitation and a pair of adjacent vertices with $q=\pm 2$ (e.g., a $z$=4 vertex with 3-in, 1-out and a neighboring $z$=2 vertex with 0-in, 2-out).  Subsequent flips of neighboring islands cause the initial dipole excitation to fractionalize into effective magnetic charges that separate and move along the 1D chain, leaving behind a string of flipped islands.  Crucially, exactly along the $iii \leftrightarrow v$ border this motion has no net cost in energy, and spin excitations can therefore readily fractionalize into quasiparticles that diffuse independently along the spin chain. The mean quasiparticle lifetime becomes very long, resulting in their rapid proliferation and formation of a regime rich in fractionalized excitations. Since the kinetics of these quasiparticles is necessarily associated with flipping of islands, this generates a telltale spike in the equilibrium noise, precisely as observed.

Similarly, Figs. 4c,d shows that noise in band B arises from the emergence and kinetics of mobile charges along the short 4-island ribs of the backbones.  Figure 4e shows a map of the calculated density of fractionalized excitations (i.e., quasiparticles separated by more than one lattice spacing). Bands A and B of the measured noise maps therefore correspond to new regimes of mobile spin excitations, stabilized by the interplay between Zeeman and interaction energies. 

Finally, note that the deconstructed noise maps in Figs. 3e,h confirm that fluctuations at $B$=0 arise solely from the staircases, which at zero field are vertex-frustrated, disordered, and extensively degenerate \cite{Morrison:2013, Gilbert:2016}. In contrast, Figs. 3(c,d,f,g) confirm that the backbones do not fluctuate at $B$=0, in line with the known stable type-I order of the $z$=4 vertices \cite{Morrison:2013, Gilbert:2016}. However, the small annular region of noise that arises from the backbone spins reveals where the $z$=4 vertices transition from type-I to stable type-II order as $|B_{x,y}|$ increases (see Supplemental Material). As a final point of interest, we note that although the horizontal backbone islands in Fig. 3c and 3d both generate annular regions of noise, the \textit{total} noise from all horizontal islands does not (see Fig. 3a). This is because, as discussed in the Supplemental Material, fluctuations of these island subsets are temporally \textit{anti}correlated, giving minimal contribution to the \textit{net} $(\delta m_x)^2$, in agreement with experiment. In contrast, fluctuations of the different vertical spins in Figs. 3f,g are temporally correlated, giving a large  total $(\delta m_y)^2$ in Fig. 3b, again in agreement with experiment and further demonstrating the utility of noise decomposition in ASIs. 

We have shown that tetris ice can host a rich variety of field-induced magnetic states. Comparing the configuration diagram (Fig.~2) and the experimental noise maps (Fig. 1) demonstrates that noise studies are a potent tool to directly detect and disentangle these states, where the critical boundaries are characterized by high kinetic activity. Crucially, certain noise signatures herald collective behaviors corresponding to the diffusion and proliferation of fractionalized magnetic quasiparticles along specific spin chains in the lattice. The fact that these phenomena are field-tunable may open the door to future functionalities in sensing or beyond-Turing computation \cite{Arava:2018, Caravelli:2020, Heyderman:2022, Gartside:2022}. Future studies will concentrate on the detailed nature of the noise spectra, and establish relations between the non-triviality of the disorder of certain phases, tied to their collective behavior, and the `color' of the noise spectrum \cite{Goryca:2021, Hallen:2022, Klyuev:2017, Nisoli:EPL2021} at larger frequencies. 

We acknowledge support from the Los Alamos LDRD program, the US Department of Energy (DOE) Quantum Science Center, the National Science Foundation (NSF) DMR-2103711, and the US DOE Office of Basic Energy Sciences, Materials Sciences and Engineering Division under Grant No. DE-SC0020162. The NHMFL is supported by NSF DMR-1644779, the State of Florida, and the US DOE.  Research at the University of Warsaw leading to these results has received funding from the Norwegian Financial Mechanism 2014-2021 under Grant No. 2020/37/K/ST3/03656 and from the Polish National Agency for Academic Exchange within Polish Returns program under Grant No. PPN/PPO/2020/1/00030. 


\newpage
\renewcommand{\thefigure}{S\arabic{figure}}
\setcounter{figure}{0}
\section{Supplemental Material}
\subsection{Configurations of tetris ice in small magnetic fields near B $\approx$ 0} 

Figure S1a shows a calculated zero-temperature magnetization map of tetris ice, focusing on the region of small applied in-plane magnetic fields $B_x$ and $B_y$. Dotted lines denote the boundaries between different stable magnetic configurations. For easier comprehension of the map, the net magnetization can be decomposed into the magnetization of staircase and backbone sublattices, shown in Figs. S1b and S1c, respectively, for which moment configurations are also explicitly shown. For the staircase sublattice, the four regions in the corners of the map correspond to fully polarized configurations, where all islands are magnetized most closely along the direction of the external magnetic field. The two central regions closest to $B=0$ reflect the situation where the phenomenon of vertex frustration prevents all the staircase vertices from achieving their lowest-energy state and half of the $z=3$ vertices remain ``unhappy''. In the case of the backbone sublattice (Fig. S1c), the central grey region indicates where all $z$=4 vertices have type-I configuration (two spins in, two spins out, with no net magnetization), which is surrounded by four stable phases where all $z$=4 vertices have type-II configuration (two spins in, two spins out, but with a net diagonal magnetization). Noise along the annular perimeter of this central region results from stochastic fluctuations within the backbones between type-I and type-II order. 

\begin{figure*}[h]
\includegraphics[width=1.9\columnwidth]{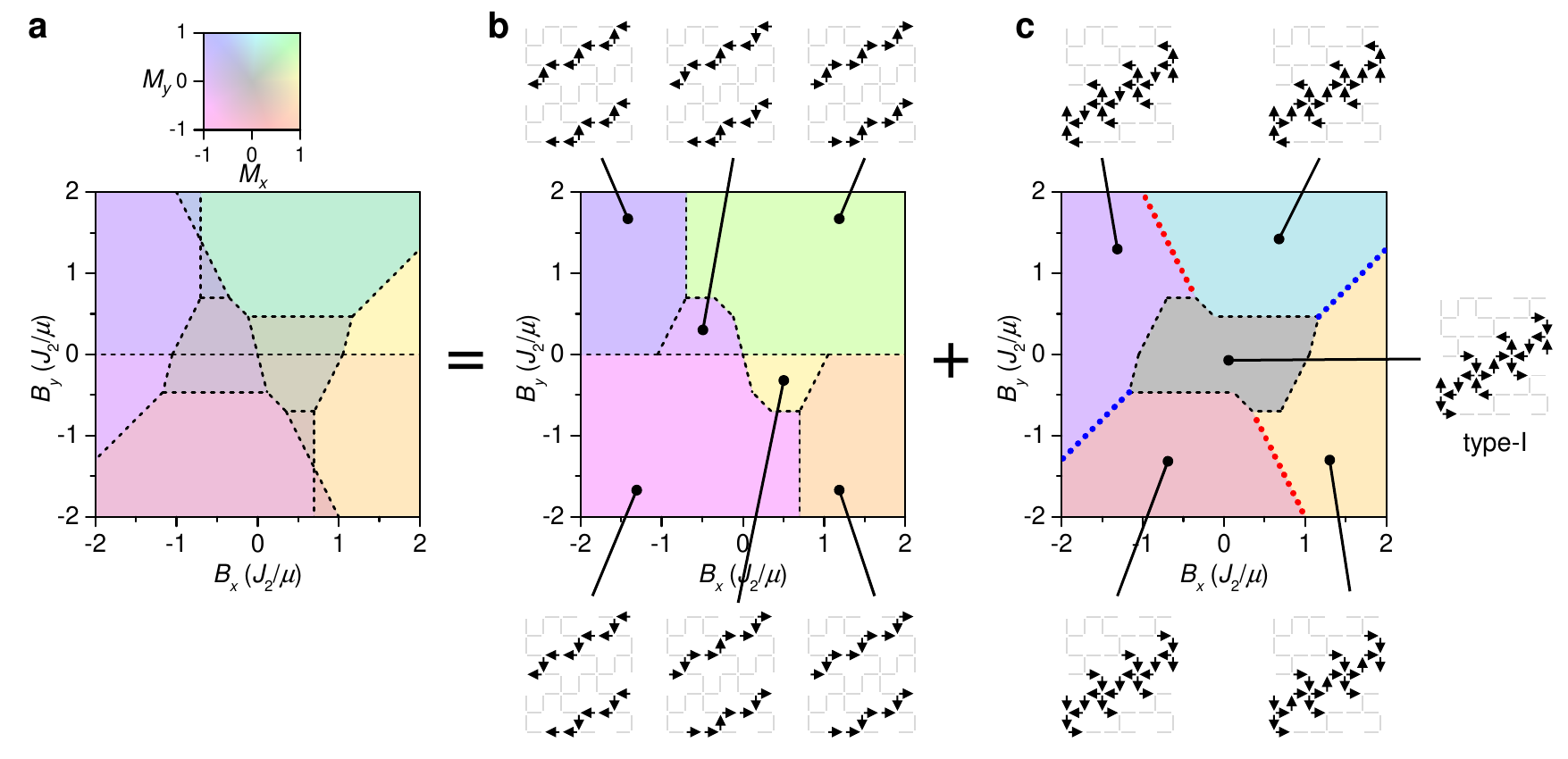}
\caption{a) Calculated zero-temperature magnetization map of tetris ice. The color and brightness indicate the direction and magnitude of the magnetization \textbf{M}. Dotted black lines denote the boundaries between different stable magnetic configurations. The net magnetization can be decomposed into the magnetization of staircase and backbone sublattices, shown in panels b) and c), respectively, for which moment configurations are explicitly shown. Inside the central grey region of panel c), the $z=4$ vertices in the backbones exhibit stable type-I order.  Outside of this region they exhibit type-II order. The red and blue dotted lines denote where increased noise is observed due to the proliferation of fractionalized magnetic excitations, discussed in the main text. The red lines follow $B_{y}=-2B_{x}$. The blue lines follow $B_{y}=B_{x}-\left(J_1-J_2\right)/\mu$ for $B_{x}>0$, and $B_{y}=B_{x}+\left(J_1-J_2\right)/\mu$ for $B_{x}<0$.}
\end{figure*}


\subsection{Mapping out temporal correlations between the noise from different subsets of islands} 

Figure S2 shows an example of the detailed insight that Monte Carlo (MC) simulations can provide into the temporal \textit{correlations} between the noise from different subsets of islands within a magnetic lattice. Figure S2a shows calculated correlations between fluctuations of different subsets of horizontal islands within the backbones (pink and green islands in the upper panel) vs. magnetic field. The correlations in the map shown in the bottom panel are calculated as $C = \left<\delta m_{\mathrm{P}}\delta m_{\mathrm{G}}\right>$, where $\delta m_{\mathrm{P}}$ and $\delta m_{\mathrm{G}}$ denote fluctuations of the pink and green island subsets, respectively. Red colors indicate positive temporal correlation; i.e., stochastic spin-flips of (say) pink islands are accompanied by flips of the green islands in the same direction within the same MC step. Blue colors indicate temporal \emph{anti}correlation; i.e., flips of pink islands in a given direction are accompanied by flips of green islands in the opposite direction, within the same MC step. 

Figure S2a shows correlations between two different subsets of horizontal islands within tetris' backbones.  The pink and green islands are those shown in Figs. 3c and 3d of the main text (respectively), where it was shown that each subset, considered alone, generates an annular region of noise surrounding $B\approx 0$. However, in combination they do not (see Fig. 3a). This is because fluctuations of the pink and green islands are strongly anticorrelated along the upper and lower horizontal borders of the annular region.  That is, when (say) a pink island flips, this rapidly drives a likely flip of the nearby green island in the opposite direction (because they are coupled via the connecting vertical island), leading to no net change in total magnetization in the horizontal direction. 

Conversely, Fig. S2b shows that the two indicated subsets of \emph{vertical} islands within the backbones generate noise that is strongly positively correlated along the upper and lower borders of the annular region. As a result, these horizontal sections of annular noise are clearly visible in the total noise from the vertical islands shown in Fig. 3b, in agreement with the experimental data. Interestingly, even though noise from these pink and green subsets in Fig. S2b is \emph{anti}correlated along the approximately vertical borders of the annular region, those sections are still visible in the total noise shown in Fig. 3b. This is because the number of the pink islands in Fig. S2b is twice the number of the green islands (which is not the case for the horizontal islands in Fig. S2a). As a result, the noise generated by the two island subsets does not cancel out completely, and the full annular region of noise is still visible in Fig. 3b. This provides a fine example of how the deconstruction of the noise into contributions from, and correlations between, different subsets of moments is a powerful tool in the analysis of the kinetics and coupling of complex magnets.


\begin{figure*}[t]

\includegraphics[width=1.1\columnwidth]{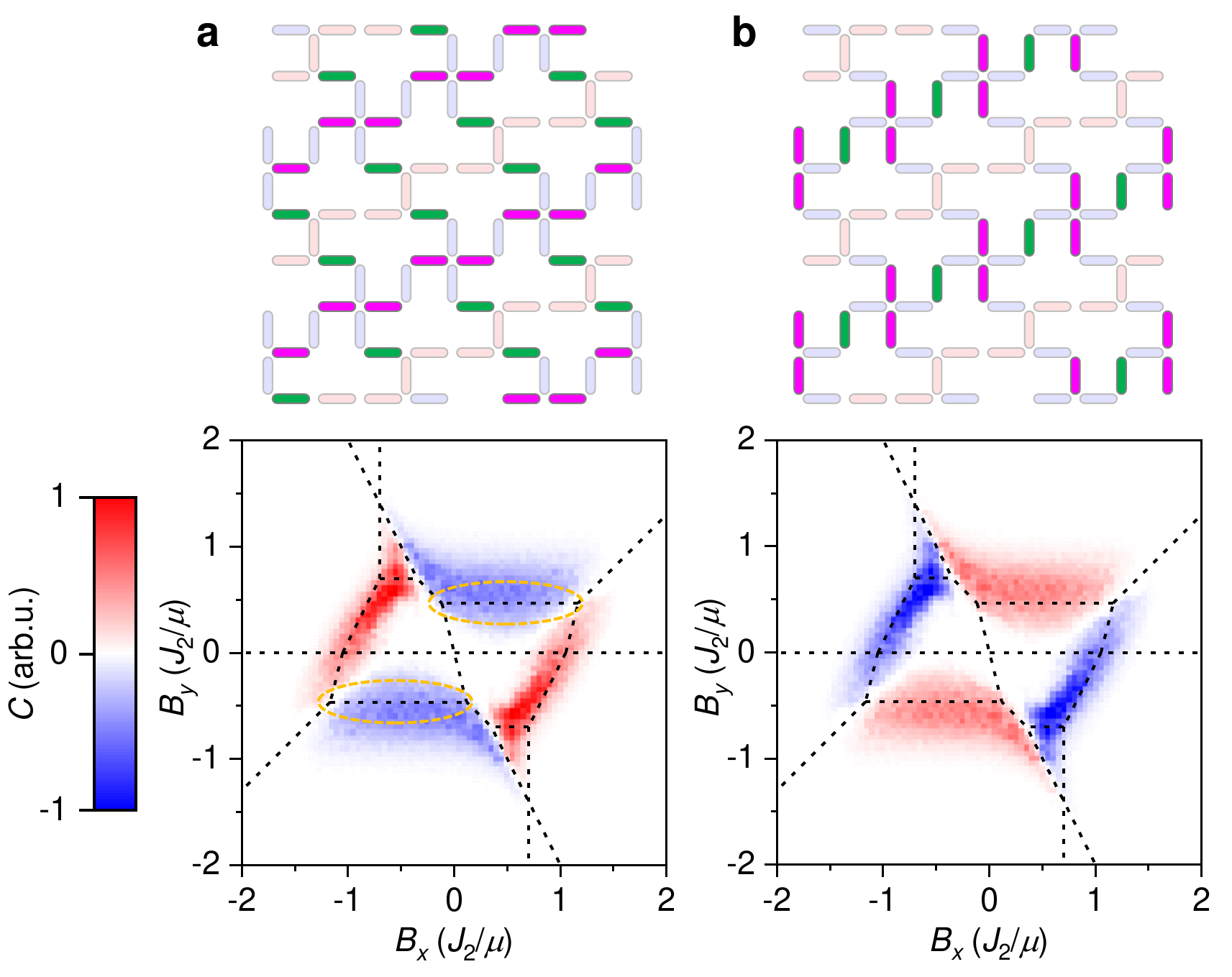}
\caption{a) Calculated temporal correlations $C$ between fluctuations of the pink and green subsets of horizontal islands (both subsets are components of tetris' backbones), vs. magnetic field. Red colors denote positive correlation (flips of pink and green islands tend to occur in the same direction, within a MC step), while blue colors denote \emph{anti}correlation (pink and green islands tend to flip in \textit{opposite} directions, within a MC step). Strong anticorrelations along the upper and lower boundaries of the central annular region indicates that noise from these two subsets largely cancels out, and therefore does not appear in the \textit{total} noise from \textit{all} the horizontal islands (see Fig. 3a of the main text). b) Same, but for temporal correlations between fluctuations from different subsets of vertical islands within the backbones.}
\end{figure*}

\end{document}